\documentstyle[epsf,rotate,preprint,prb,aps]{revtex}
\tightenlines
\def\inseps#1#2{\def\epsfsize##1##2{#2##1} \centerline{\epsfbox{#1}}}
\input{mssymb.tex}
\def\opeq{\mathop{=}}
\def\oppropto{\mathop{\propto}} 
\def\opsimeq{\mathop{\simeq}}

\def\oplimit{\mathop{ {\rm lim }}} 

\title{Copolymer at a selective interface and two dimensional wetting:\\
a grand canonical approach}
\author{C. Monthus \footnote{On leave from LPTMS, Bat.100,
91406 Orsay, France}, T. Garel and H. Orland\\
Service de Physique Th\'eorique\\
CE-Saclay, 91191 Gif-sur-Yvette Cedex, France}

\date{\today}

\def\be{\begin{equation}}
\def\bea{\begin{eqnarray}}
\def\ee{\end{equation}} 
\def\eea{\end{eqnarray}} 
\begin{document}
\maketitle
\vskip 5mm
\begin{abstract}
We consider two different problems involving the localization of a
single polymer chain: (i) a periodic $AB$ copolymer
at a selective fluid-fluid interface, with the upper (resp. lower)
fluid attracting $A$ (resp. $B$) monomers (ii) a homopolymer chain
attracted to a hard wall (wetting). Self avoidance is neglected in both
models, which enables us to study their localization transition in
a grand canonical approach. We 
recover the results obtained in previous studies via transfer matrix
methods. Moreover, we calculate in this way the loop length
distribution functions in the localized phase. Some finite size 
effects are also determined and tested numerically.
\end{abstract}
\vskip 5mm
\noindent\mbox{Submitted for publication to: ``Eur. Phys. J.  B''} \hfill 
\mbox{Saclay, SPhT/00-037}\\ \noindent \mbox{ }\\ 
\vskip 1cm\noindent \mbox{PACS: 05.70 Np; 61.41.+e; 64.70.-p; 68.45
Gd} \newpage 

\section{Introduction}
\label{sec:intro}
The behaviour of polymers at interfaces is currently a subject of
great activity \cite {Fle_Coh_Sch,Deb_Tur}. In this paper, we focus
our interest on two problems, namely (i) the adsorption of a periodic
copolymer chain on a selective fluid-fluid interface (ii) the
adsorption of a homopolymer chain on a hard wall (wetting). These
problems have been previously solved for one dimensional geometries, provided
certain assumptions are fulfilled (self avoidance is neglected,
wetting is considered in the SOS approximation,...). A quick guide to
the recent litterature can be found in references
\cite{For_Luc_Nie_Orl,Der_Hak_Van,Gro_Izr_Nec,Nec_Zha,Som_Da1,Som_Da2}
and references therein. 
Here, we will tackle these models, with the same assumptions as above,
through the use of a grand canonical method. This
method was initiated in the context of the DNA helix-coil transition
\cite{Lif_Zim,Lif,Pol_Sch1,Pol_Sch2}
 and was recently used in the
context of RNA folding \cite{Bun_Hwa}. We will recover previously
known results about models (i) and (ii). Moreover, this approach
enables us to obtain the loop length distribution function in the
localized regime. We also find in this way some analytic expressions
of finite size corrections at the critical temperature, which are in
agreement with numerical transfer matrix calculations.
The lay out of the paper is as follows. Section
\ref{sec:periodic} deals with the periodic $AB$ copolymer chain at
a selective interface. In section \ref{sec:wetting}, we consider the
wetting of a homogeneous wall by a homopolymer chain.

\section{Periodic copolymer chain at a selective interface}
\label{sec:periodic}

\subsection{Introduction}
We consider a periodic $AB$ copolymer chain of $2N$ monomers at a
selective fluid-fluid interface, located at $z_0={1 \over 2}$. The
upper (resp. lower) fluid 
is a good solvent for $A$-type (resp. $B$-type) monomers, and will
herefrom be called fluid $\alpha$ (resp. fluid $\beta$). This
situation is modelled through the Hamiltonian
\begin{equation}
\label{hamil}
{\cal H}_{2N}=-\sum_{i=1}^{2N} q_i \ {\rm sgn} [z(i)-\frac{1}{2}]
\end{equation}
where ${\rm sgn}(u)$ denotes the sign function, and the ``charges'' $q_i$ of
the periodic chain are chosen as $q_{2n+1}=q_A \ >0 $
and $q_{2n}=-q_B \ <0$, together with $q_A \ge q_B$. In equation
(\ref{hamil}), $z(i)$ denotes the 
position of monomer $i$ of the chain, which we represent
as an (unidimensional) random walk (RW) with the properties
$z(i)=..,-2,-1,0,+1,+2,..$, and $z(i+1)-z(i)=\pm 1$. We emphasize once 
more that self avoidance is neglected in this model.

The partition function then reads
\begin{eqnarray}
\label{partit}
Z_{2N}= \sum_{(RW)} e^{-\beta {\cal H}_{2N}}
\end{eqnarray}
where $\beta ={1 \over T}$ and the sum runs over all random walks (RW)
starting with an $A$ monomer 
at $z(i=1)=1$. The thermodynamical properties of  this model can be
obtained through the use of transfer matrix methods
\cite{Gro_Izr_Nec}, as given in Appendix ${\rm A}$. Here, we follow
references \cite{Lif_Zim,Lif,Pol_Sch1,Pol_Sch2}, and  consider a
particular chain configuration as a collection of non interacting
loops. This class of problems is solved in a grand canonical approach, 
by defining a grand canonical partition function
\begin{eqnarray}
\label{grandcano}
Z(K)=\sum_{N=1}^{+\infty} K^{ N} Z_{2N}
\end{eqnarray}
and noting that the radius of convergence $K_*(\beta)$ of the series
$Z(K)$ gives the free energy per monomer in the thermodynamic limit
\begin{eqnarray}
\label{free ener}
f= \frac{ \log(K_*(\beta)) }{ 2 \beta}
\end{eqnarray} 

Moreover, one can easily calculate in this way quantities such as the
loop length distribution function, which are not accessible through
the transfer matrix approach.

\subsection{Calculation of the grand canonical partition function}
\label{grand cano}

 The partition function $Z_{2N}$ for a chain of $(2N)$ monomers
can then be decomposed according to the position $z(2N)$ of the end
point of the chain. For simplicity, we set $z(2N)=2n$, and write
\begin{eqnarray}
\label{decompo}
Z_{2N}=\sum_{n=-\infty}^{+\infty} Z_{2N} (2n)
\end{eqnarray}
 Given the expression of the Hamiltonian (\ref{hamil}), the chain
configurations can be further decomposed into loops. A preliminary
remark is as follows: if ${\cal N}(l,z)$ denotes the number of random
walks of $l$ steps going from $z=0$ to $z \geq 0$ in the presence of
an absorbing barrier at $z=0^-$, the image method \cite{Feller} yields
\begin{eqnarray}
&& {\cal N}(2l,2p) = C_{2l}^{l+p} - C_{2l}^{l+p+1} \\
&& {\cal N}(2l-1,2p-1) = C_{2l-1}^{l+p-1} - C_{2l-1}^{l+p}
\end{eqnarray}
depending on the parity of $(l,z)$.

Let us temporarily suppose that the chain end is in fluid $\alpha$
$(2n \ge 2)$. The first loop, located by hypothesis in fluid
$\alpha$ contains $(2l_1-1)$ monomers, namely $l_1$ A monomers and 
$(l_1-1)$ B monomers. The second loop, located in fluid $\beta$,
contains $(2l_2-1)$ monomers, namely $(l_2-1)$ A monomers and $l_2$ B
monomers.... The last loop is denoted by $l_{2p}$. Finally, one must
also specify the number $2l$  of monomers between the interface and
the chain end.  We may then write 
\begin{eqnarray}
\label{complique}
Z_{2N}(2n)
&& =  \sum_{p=0}^{+\infty} \sum_{l_1=1}^{\infty} \sum_{l_2=1}^{\infty}
... \sum_{l_{2p}=1}^{\infty} \sum_{l=1}^{\infty} \delta(2N-(
\sum_{i=1}^{2p} (2 l_i-1)+2l) ) 
e^{\beta q_B + \beta (q_A-q_B) l_1} {\cal N}(2l_1-2,0) \nonumber
\\
&& e^{\beta q_A - \beta (q_A-q_B) l_2}  {\cal N}(2l_2-2,0) .....
e^{\beta q_A - \beta (q_A-q_B) l_{2p}}  {\cal N}(2l_{2p}-2,0) \nonumber\\
&&e^{ \beta (q_A-q_B) l} {\cal N}(2l-1,2n-1)
\end{eqnarray}

Plugging equation (\ref{complique}) into equations
(\ref{grandcano}) and (\ref{decompo}), we get 
\begin{equation}
\label{grandcano1}
Z(K)=\sum_{n=-\infty}^{+\infty} Z (K,2n)
\end{equation}
where
\begin{eqnarray}
\label{grandcano2}
Z(K,2n) =&& \sum_{N=1}^{+\infty} K^{ N} Z_{2N}(2n) \nonumber\\
=&& \frac{1}{1-e^{\beta (q_A+q_B)} B(K^{1/2} e^{\beta q_0} )  B(K^{1/2} e^{-\beta
q_0} ) }   \sum_{l=1}^{\infty} ( K e^{2 \beta q_0})^{ l}   {\cal
N}(2l-1,2n-1)  
\end{eqnarray}
with 
\begin{eqnarray}
q_0=\frac{q_A-q_B}{2}
\end{eqnarray}
and
\begin{eqnarray}
B(y)=\sum_{l=1}^{+\infty}  y^{2l-1} {\cal N}(2l-2,0)= \frac{1}{2y} \left(1- \sqrt{1-4y^2}\right)
\end{eqnarray}
We finally get, for ($ 2n \ge 2$)
\begin{eqnarray}
\label{grandcano3}
Z(K,2n)  = \sum_{N=1}^{+\infty} K^{ N} Z_{2N}(2n)  
 = \frac{1}{1-e^{\beta (q_A+q_B)} B(K^{1/2} e^{\beta q_0} )  B(K^{1/2} e^{-\beta q_0} ) }  
\left( \frac{B(K^{1/2} e^{\beta q_0})}{K^{1/2} e^{\beta q_0}}-1 \right)^n
\end{eqnarray}

Similarly, if the chain end is in fluid $\beta$ ($2n \leq 0$), we find
\begin{eqnarray}
\label{grandcano4}
Z(K,2n) = \sum_{N=1}^{+\infty} K^{ N} Z_{2N}(2n)  
=\frac{ e^{\beta (q_A+q_B)} B(K^{1/2} e^{\beta q_0} ) B(K^{1/2} e^{-\beta q_0} )}
{1-e^{\beta (q_A+q_B)} B(K^{1/2} e^{\beta q_0} )  B(K^{1/2} e^{-\beta q_0} ) }
 \left( \frac{B(K^{1/2} e^{-\beta q_0})}{K^{1/2} e^{-\beta q_0}}-1 \right)^{\vert n \vert} 
\end{eqnarray}

\subsection{Thermodynamics}

We finally get from (\ref{grandcano3}) and  (\ref{grandcano4})
\begin{eqnarray}
Z(K) =  \sum_{n=-\infty}^{+\infty} Z (K,2n)  = 
\frac{  r_+( K^{1/2} e^{\beta q_0}) + e^{\beta (q_A+q_B)} B(K^{1/2} e^{\beta q_0} )
 r_- (K^{1/2} e^{-\beta q_0})}
{1-e^{\beta (q_A+q_B)} B(K^{1/2} e^{\beta q_0} )  B(K^{1/2} e^{-\beta q_0} ) }
\label{resum}
\end{eqnarray}
where
\begin{eqnarray}
&& r_+(y)= \frac{\frac{B(y)}{y}-1}{2-\frac{B(y)}{y}} = \frac{1}{2} \left( \frac{1}{\sqrt{1- 4y^2}}-1 \right)  \\
&& r_-(y)= \frac{B(y)}{2-\frac{B(y)}{y}}
= \frac{y}{\sqrt{1- 4y^2}}
\end{eqnarray}

The radius of convergence $K_*(\beta)$ of the grand canonical
partition function $Z(K)$ is
either given by the singularity of the numerator of equation
(\ref{resum}), or by the singularity of the denominator. The former
corresponds to the value 
\begin{eqnarray}
K_{deloc}(\beta)=\frac{1}{4} e^{ - 2 \beta q_0 }
\end{eqnarray}
The latter comes from the resummation of the loops, and is given by
\begin{eqnarray}
\label{kstar}
K_*(\beta)=\frac{\sinh(\beta q_A) \sinh(\beta q_B) }
{\sinh^2 \beta (q_A+q_B))}
\end{eqnarray}

If the solution $K_*(\beta)$ is smaller than $K_{deloc}$ then the
chain is localized. This actually happens for $\beta>\beta_c$
(i.e. $T<T_c$), where $\beta_c$ is the solution of 
\begin{eqnarray}
e^{2 \beta_c q_B} + e^{-2 \beta_c q_A}-2=0
\label{eqtc}
\end{eqnarray}
In particular, in the limit $q_0=(q_A-q_B)/2 \to 0$, the critical
temperature $T_c$  diverges as
\begin{eqnarray}
T_c \opsimeq_{q_{0} \to 0} \frac{q_A^2}{q_0}
\end{eqnarray}
Note that the scaling of $T_c$ with $q_0$ differs from equation (10)
of reference \cite{Som_Da1}, since the model used in this
reference allows the monomers to sit at the interface, in contrast to
equation (\ref{hamil}). The physics is nevertheless the
same: a symmetric copolymer chain ($q_A=q_B$) is always localized at
the interface.

Since the radius of convergence $K_*(\beta)$ and free energy per
monomer are linked by equation (\ref{free ener}), we get
\begin{eqnarray}
f(T) 
= \frac{1}{\beta} \ln
\frac{ \sqrt{\sinh(\beta q_A) \sinh(\beta q_B) } } { \sinh\beta (q_A+q_B)}
\label{freesym}
\end{eqnarray}
for $T \le T_c$, which in turn yields
\begin{eqnarray}
\label{critic}
f(T)  \opsimeq_{T \to T_c} f_{deloc}(T_c) - C (T_c-T)^2
\end{eqnarray}
where $f_{deloc}(T_c)=-T_c \ln 2 -q_0$ is the free energy of the
delocalized phase (since we have chosen $q_0 >0$, the chain is then in
fluid $\alpha$). Equation (\ref{critic}) corresponds to a critical
exponent $\alpha=0$. The jump in the 
specific heat  per monomer is given by
\begin{eqnarray}
\label{specifinf}
&& C_{\infty}(T_c^-)=\left(  \frac{(2 e^{2  \beta_c q_A}-1)
\ln (2-e^{-2 \beta_c q_A}) - 2 \beta_c q_A }
{2 (e^{2  \beta_c q_A}-1) } 
\right)^2  \\
&& C_{\infty}(T_c^+)=0
 \end{eqnarray}

\subsection{Density profile}
We denote the normalized density profile by $\rho(2n)$ and write
\begin{eqnarray}
\label{rho}
\rho(2n) = \oplimit_{N \to \infty} \frac{Z_{2N}(2n)}{ Z_{2N} }
\end{eqnarray}
Inverting equations (\ref{grandcano3}) and (\ref{grandcano4}),  we
have
\begin{eqnarray}
\label{inversion1}
 Z_{2N}(2n) = \oint_{\hbox{ circle around $0$}} \frac{dK}{2 i \pi} \frac{Z(K,2n)}{K^{ N+1}}
\end{eqnarray}
After deformation of the contour, the dominant contribution for large $N$
comes from the pole at $K_*(\beta)$ (see equation (\ref{kstar})), 
and reads
\begin{eqnarray}
 Z_{2N}(2n) && \opsimeq_{N \hbox{large}}
-   \oint_{\hbox{ circle around $K_*$}} \frac{dK}{2 i \pi} \frac{Z(K,2n)}{K^{N+1}}
\end{eqnarray}
Summing over $n$ leads to
\begin{eqnarray}
 Z_{2N} && \opsimeq_{N \hbox{large}}
-  \oint_{\hbox{ circle around $K_*$}} \frac{dK}{2 i \pi} \frac{Z(K)}{K^{ N+1}}
\end{eqnarray}

The density profile $\rho(2n)$ is then obtained as 
\begin{eqnarray}
\rho(2n) =
\frac{1}{  r_+( K_*^{1/2} e^{\beta q_0}) + e^{\beta (q_A+q_B) } B(K_*^{1/2} e^{\beta q_0} )
 r_- (K_*^{1/2} e^{-\beta q_0}) }
\left( \frac{B(K_*^{1/2} e^{\beta q_0})}{K_*^{1/2} e^{\beta q_0}}-1 \right)^n  
\eea
for $n\geq 1$ and
\bea
\rho(2n) =
\frac{ e^{\beta (q_A+q_B)} B(K_*^{1/2} e^{\beta q_0} ) B(K_*^{1/2} e^{-\beta q_0} )}
{  r_+( K_*^{1/2} e^{\beta q_0}) + e^{\beta (q_A+q_B)} B(K_*^{1/2} e^{\beta q_0} )
 r_- (K_*^{1/2} e^{-\beta q_0}) }
\left( \frac{B(K_*^{1/2} e^{-\beta q_0})}
{K_*^{1/2} e^{-\beta q_0}}-1 \right)^{\vert n \vert}
\end{eqnarray}
for $n \leq 0$.
Inserting the value of $K_*(\beta)$, we get
\begin{eqnarray}
&& \rho(2n)=  \frac{ ( e^{2 \beta q_B}+ e^{-2 \beta q_A} -2 ) ( e^{2 \beta q_A}+ e^{-2 \beta q_B} -2 )}
{ (e^{2 \beta (q_A+q_B)} -1) (1-e^{-2 \beta q_A}) (1-e^{-2 \beta q_B}) }
\left( \frac{1-e^{-2 \beta q_A}}{e^{2 \beta q_B}-1}\right)^{n}
\ \ \ \ \hbox{for}\ \  n \geq 1  \\
&& \rho(2n)= \frac{ ( e^{2 \beta q_B}+ e^{-2 \beta q_A} -2 ) ( e^{2 \beta q_A}+ e^{-2 \beta q_B} -2 )}
{ (e^{2 \beta (q_A+q_B)} -1) (1-e^{-2 \beta q_A}) (1-e^{-2 \beta q_B}) }
\left( \frac{1-e^{-2 \beta q_B}}{e^{2 \beta q_A}-1} \right)^{-n} 
\ \ \ \ \hbox{for} \ \ n \leq 0 
\label{pstar}
\end{eqnarray}

i.e. the density around the interface decays exponentially from the interface
with characteristic lengths $\xi_{\alpha}$ (resp. $\xi_{\beta}$) in 
fluid $\alpha$ (resp.  fluid $\beta$), with
\begin{eqnarray}
\xi_{\alpha}(T) = && \frac{2}{\ln \left( \frac {e^{2 \beta q_B}-1} {1-e^{-2 \beta q_A}} \right)}\\
\xi_{\beta}(T)= && \frac{2}{ \ln \left(\frac{e^{2 \beta q_A}-1}{1-e^{-2 \beta q_B}} \right)}
\end{eqnarray}
Close to $T_c$, we get
\begin{eqnarray}
\xi_{\alpha}(T) \oppropto_{T \to T_c} \frac{1}{T_c-T}
\end{eqnarray}
and
\begin{eqnarray}
\xi_{\beta}(T_c) = \frac{T_c}{(q_A+q_B) }
\end{eqnarray}

The order parameter of the transition can be chosen as the probability 
${\cal M}_{\beta}$
to be in fluid $\beta$ and vanishes linearly at the transition:
\begin{eqnarray}
\label{ordparam}
{\cal M}_{\beta}=\sum_{n=-\infty}^0 \rho(2n) \oppropto_{T \to T_c} (T_c-T)
\end{eqnarray}

\subsection{Probability distribution of loop lengths}
\label{subsec:probadis}

Contrary to the direct matrix transfer approach, the grand canonical
approach allows one to obtain informations 
on the statistical properties of loops in each fluid.
Without loss of generality, we may consider the partition function $Z_{2N}(2n=0)$
 where the chain of length $2N$
is attached at the interface at both ends,
and decompose it into 
\begin{eqnarray}
Z_{2N}(2n=0)=\sum_{l_1=1}^{\infty} \sum_{l_2=1}^{\infty} Z_{2N}(2n=0,2 l_1-1,2l_2-1)
\end{eqnarray}
where $Z_{2N}(2n=0, 2 l_1-1,2l_2-1)$
represents the partition function with the constraints that the first
loop in fluid $\alpha$
contains exactly $(2 l_1-1)$ monomers and that the first loop in fluid 
$\beta$ contains exactly $(2 l_2-1)$ monomers.

 We again define generating functions by
\begin{eqnarray}
{\hat Z} (K,s_1,s_2) = \sum_{N=1}^{\infty} \sum_{l_1=1}^{\infty} \sum_{l_2=1}^{\infty} 
K^{N} e^{-s_1 (2 l_1-1) -s_2 (2 l_2-1)} Z_{2N}(2n=0,2 l_1-1,2l_2-1)
\end{eqnarray}

The generalization of equation (\ref{grandcano4}) to the case $(s_1,
s_2) \neq (0,0)$ then yields
\begin{eqnarray}
\label{general1}
{\hat Z} (K,s_1,s_2) = \frac{{e^{\beta (q_A+q_B)} B(K^{1/2} e^{\beta q_0} e^{-s_1} )  B(K^{1/2} e^{-\beta q_0} e^{-s_2}) }   }{1-e^{\beta (q_A+q_B)} B(K^{1/2} e^{\beta q_0} )  B(K^{1/2} e^{-\beta q_0} ) }
\end{eqnarray}

Again, by inversion and deformation of the contour in the complex plane,
we get the asymptotic behavior for large $N$
\begin{eqnarray}
{\hat Z}_{2N}(s_1,s_2)  =&& \oint_{\hbox{ circle around $0$}}
\frac{dK}{2 i \pi}  \frac{{\hat Z} (K,s_1,s_2)}{K^{ N+1}} \nonumber\\
&&\opsimeq_{N \hbox{large}}-  \oint_{\hbox{circle around $K_*$}}
\frac{dK}{2 i \pi}  
\frac{{\hat Z} (K,s_1,s_2)}{K^{ N+1}}
\end{eqnarray}
The generating functions ${\hat P}_{\alpha}$ and ${\hat P}_{\beta}$ of
the loop length distribution functions in the two fluids are given by
\begin{eqnarray}
 && {\hat P}_{\alpha}(s) \equiv \sum_{l_1=1}^{\infty} e^{-s (2 l_1-1)} P_{\alpha}(2 l_1-1)
= \oplimit_{N \to \infty} \left(  \frac{ {\hat Z}_{2N}(s_1=s,s_2=0)}{
{\hat Z}_{2N}(s_1=0,s_2=0)}  \right) 
\\
 && {\hat P}_{\beta}(s) \equiv \sum_{l_2=1}^{\infty} e^{-s (2 l_2-1)} P_{\beta}(2 l_2-1)
= \oplimit_{N \to \infty} \left( \frac{ {\hat Z}_{2N}(s_1=0,s_2=s)}{ {\hat Z}_{2N}(s_1=0,s_2=0)}  \right)
\end{eqnarray}

More explicitely, we have
\begin{eqnarray}
\label{hat2}
 && {\hat P}_{\alpha}(s)= \frac{ B(K_*^{1/2} e^{\beta q_0} e^{-s} ) }{
 B(K_*^{1/2} e^{\beta q_0} ) } = e^s \frac { 1-\sqrt{1-e^{- \omega(\beta) -2 s} } }
 { 1-\sqrt{1-e^{- \omega(\beta) } } } \\
 && {\hat P}_{\beta}(s)= \frac{ B(K_*^{1/2} e^{-\beta q_0} e^{-s} ) }
{ B(K_*^{1/2} e^{\beta q_0} ) } = e^s \frac{ 1-\sqrt{1-e^{- \omega(\beta) -4\beta q_0 -2 s} } }
 { 1-\sqrt{1-e^{- \omega(\beta) -4\beta q_0 } } }
\label{hat1}
\end{eqnarray}
with
\begin{eqnarray}
 \omega(\beta)  = -2 \beta q_0 - 2 \ln 2 -  \ln K_* = 2 \beta (f_{deloc}-f)
=\ln \left( 
\frac { ( e^{2 \beta (q_A+q_B) } -1 )^2 }
{ 4  e^{2 \beta q_A} (e^{2 \beta q_A}-1  ) ( e^{2 \beta q_B}  -1 ) }
 \right)
\label{omega}
 \end{eqnarray}

We note that equations (\ref{hat2}) and (\ref{hat1}) exhibit several
loop size scales, that we consider below.

At the critical temperature ($T=T_c$), the distribution of loop lengths
$(l_{\alpha})$ in fluid $\alpha$ is simply the loop distribution of a
free random walk
\begin{eqnarray}
{\hat P}_{\alpha}(s) \opeq_{T = T_c} e^s 
 \left[ 1-\sqrt{1-e^{ -2 s} } \right] = 1- \sqrt{2s}+ O(s)
\end{eqnarray}
yielding an algebraic decay for large $l_{\alpha}$
\begin{eqnarray}
 P_{\alpha}(l_{\alpha}) \opsimeq_{l_{\alpha} \to \infty}
\frac{1}{l_{\alpha}^{3/2}} 
\label{free}
\end{eqnarray}
A consequence of this critical distribution ($T=T_c$) is as
follows. For a chain  of length $N$, the number of loops scales as
$\sqrt{N}$, the longest ($\alpha$) loop being of order $N$.

In the critical region $T \to T_c$ where $ \omega(\beta) \sim (T_c-T)^2 $,
the moments of $(l_{\alpha})$ diverge as
\begin{eqnarray}
\overline{ l_{\alpha}^n }= \sum_{l=1}^{\infty} (2 l-1)^{n}
P_{\alpha}(2 l-1)  \oppropto_{T \to T_c} 
(T_c-T) \left( \frac{1}{ (T_c-T)^2} \right)^n  
 \end{eqnarray}
i.e. for large $l_{\alpha}$ the correct scaling variable is $\lambda=
l_{\alpha} (T_c-T)^2$, but the normalisation of the region of finite
$\lambda$ varies as $(T_c-T)$. 
This can be understood by considering the normalisation
of the region of loops bigger than a given scale $l_c \sim
\frac{1}{(T_c-T)^2}$ 
for a free random walk (\ref{free})
\begin{eqnarray}
\int_{l_c}^{\infty} dl  \frac{1}{l^{3/2}} \sim \frac{1}{l_c^{1/2}} \sim (T_c-T)
 \end{eqnarray}

In other words, the phase transition is driven by a small fraction of
large loops.

 It is of interest to note that the weight of the small $l_{\alpha}$
loops is finite, even at $T_c$.  For instance, the probability to have
 $l_{\alpha}=1$ is 
\begin{eqnarray}
P_{\alpha}(l_{\alpha}=1) = \frac{e^{2 \beta q_B}-1}{e^{2 \beta q_B}-e^{-2 \beta q_A}} 
\opsimeq_{T \to T_c} \frac{1}{2} 
 \end{eqnarray}

At zero temperature we of course recover the ground state
\begin{eqnarray}
P_{\alpha}(l_{\alpha}) \opsimeq_{T = 0} \delta_{l_{\alpha},1} 
\end{eqnarray}

In marked contrast, the critical distribution of loop lengths in the
($\beta$) fluid is given by
\begin{eqnarray}  
 {\hat P}_{\beta}(s) \opsimeq_{T = T_c} e^s
 \frac{  1-\sqrt{1-e^{ -4\beta_c q_0 -2 s}  }  }
 { 1-\sqrt{1-e^{ -4\beta_c q_0 } } }
\end{eqnarray}
leading to finite $\overline{ l_{\beta}^n }$ moments.

\subsection{Correlation function}

To further emphasize the meaning of the scale corresponding to
$\omega(\beta)$ in equation (\ref{omega}), we consider the connected
correlation function
\bea
\label{correla1}
{\cal C}(j-i)=<{\rm sgn}[z(i)-\frac{1}{2}] \
{\rm sgn}[z(j)-\frac{1}{2}]>-<{\rm sgn}[z(i)-\frac{1}{2}]> <{\rm
sgn}[z(j)-\frac{1}{2}]> 
\eea
Its generating function reads
\bea
\label{correla2}
\hat{{\cal C}}(K)=\sum_{j=i}^{\infty} K^{j-i} \  {\cal C}(j-i)
\eea
and can be expressed, via a loop decomposition, in terms of the
functions ${\hat P}_{\alpha}(s)$ and ${\hat P}_{\beta}(s)$ defined in
(\ref{hat2},\ref{hat1}), with the replacement $(s=-\ln K)$. Inverting
(\ref{correla2}), we have 
\bea
\label{correla3}
{\cal C}(j-i)=\oint_{\hbox{circle around $0$}} \frac{dK}{2 i \pi}  
\frac{{\hat {\cal C}} (K)}{K^{1+j-i}}
\eea
At large $(j-i)$ separation, the behaviour of ${\cal C}(j-i)$ is
dominated by the singularity at $K=e^{{\omega(\beta) \over 2}}$ of
${\hat P}_{\alpha}(-\ln K)$, leading to
\bea
\label{correla4}
{\cal C}(j-i) \oppropto_{(j-i) \to \infty} e^{-(j-i){\omega(\beta)
\over 2}}
\eea
This shows that $\omega(\beta)$ is the inverse of the correlation
length along the chain.

\subsection{Finite size properties}

In this section, we again consider that both chain ends are fixed at
the interface. This means in particular that one has
$Z_{2N}=Z_{2N}(0), Z(K)=Z(K,0),....$.
\subsubsection{Free energy }
In the inversion formula of equation (\ref{inversion1}), we set $2n=0$
and get
\begin{eqnarray}
 Z_{2N}(0) = \oint_{\hbox{ circle around $0$}} \frac{dK}{2 i \pi} \frac{Z(K,0)}{K^{ N+1}}
\end{eqnarray}
where $Z(K,0)$ given in (\ref{grandcano4}) presents a simple pole at
$K_*(\beta)$, and a cut on the real axis, namely
$[{K_{deloc}(\beta),+\infty}]$.

In the localized phase, and after deformation of the contour, the
dominant contribution for large $N$
comes from the pole at $ K_*(\beta)$, and
the correction is given by the leading contribution of the
cuts. Simplifying as explained above the notations, we have
\begin{eqnarray}
 Z_{2N} && \opsimeq_{N \hbox{large}}
-  \oint_{\hbox{circle around $K_*$}} \frac{dK}{2 i \pi}
\frac{Z(K)}{K^{ N+1}} 
+ O \left(\frac{ 1} {K_{deloc}^{N}}\right) \\
&& \opsimeq_{N \hbox{large}}
\frac{ 1} { K_*^{N}} \left( \rho(0)+ O \left(\frac{ K_*}
{K_{deloc}}\right)^{N} \right) 
\end{eqnarray}
where $\rho(0)$ given in (\ref{pstar}) represents the probability to
be at $2n=0$.

In the delocalized phase, the finite size properties are given by the
contributions of the cuts 
\begin{eqnarray}
 Z_{2N} && =
  \int_{0}^{+\infty}  \frac{dx}{2 i \pi} \frac{1}{K_{deloc}^{ N}
(1+x)^{N+1}} \left[ Z(K_{deloc}(1+x+io)) -
Z(K_{deloc}(1+x-io)) \right] 
\end{eqnarray}

To obtain the dominant behavior at large $N$, we will set $v= N \ln (1+x)$.

For $T>T_c$, using
\begin{eqnarray}
 Z(K_{deloc}(1+x+io)) - Z(K_{deloc}(1+x-io)) 
=\frac{2i \sqrt{ x} (1+ G(\beta) ) } {G(\beta)^2} \left(1+O(x)  \right)
\end{eqnarray}
with
\begin{eqnarray}
 G(\beta) =e^{2 \beta q_A} (1-\sqrt{1-e^{-4 \beta q_0}})-1
\label{gbeta}
\end{eqnarray}
we get the asymptotic behavior
\begin{eqnarray}
 Z_{2N} && \opsimeq_{N \hbox{large}} 
  \int_{0}^{+\infty}  \frac{d v}{2 N i \pi} \frac{1}{K_{deloc}^{ N}} e^{-v} 
\left[ Z(K_{deloc} (e^{\frac{v}{N}}+io)) -
Z(K_{deloc}(e^{\frac{v}{N}}-io)) \right] \\ 
&& \opsimeq_{N \hbox{large}} 
\frac{1}{ K_{deloc}^{N}  \sqrt{\pi} N^{3/2}} \frac{  (1+G(\beta))}{ 2
G(\beta)^2} 
\label{finitedel}
\end{eqnarray}

For $T=T_c$, $G(\beta_c)$ vanishes and we have
\begin{eqnarray}
 Z(K_{deloc}(1+x+io)) - Z(K_{deloc}(1+x-io)) 
= i \frac{2}{\sqrt{x}} \left(1+O(x)  \right)
\end{eqnarray}
leading to
\begin{eqnarray}
 Z_{2N} && \opsimeq_{N \hbox{large}} 
  \int_{0}^{+\infty}  \frac{d v}{2 N i \pi} \frac{1}{K_{deloc}^{ N}} e^{-v} 
\left[ Z(K_{deloc} (e^{\frac{v}{N}}+io))
 - Z(K_{deloc}(e^{\frac{v}{N}}-io)) \right] \\
&& \opsimeq_{N \hbox{large}} 
\frac{1}{ K_{deloc}^{N} \sqrt{\pi  N} }
\label{finitetc}
 \end{eqnarray}

This shows that the probability of presence at the interface decays 
with the length $N$ as $1/N^{1/2}$
at $T=T_c$ (\ref{finitetc}), whereas it decays as $1/N^{3/2}$ for $T > T_c$.

In the following, we will need the general scaling form which contains
(\ref{finitedel}) and ({\ref{finitetc}) as special cases 

\begin{eqnarray}
 Z_{2N}  \opsimeq_{N \hbox{large}} 
\frac{1+G(\beta)}{  K_{deloc}^{N} \sqrt{  N} H(\beta) }
\ \ q \left(\sqrt{N} \frac{G(\beta)}{\sqrt{H(\beta)}} \right)
\label{finitescaling}
 \end{eqnarray}
where the scaling function $q(x)$ reads
\begin{eqnarray}
q(x)= \frac{1}{\pi} \int_0^{+\infty} dv \frac{e^{-v} \sqrt{v}}{v+x^2}
 \end{eqnarray}
and 
\begin{eqnarray}
H(\beta)=\frac{e^{4 \beta q_B} (1-e^{-2  \beta q_A})}
{\sqrt{1-e^{-4 \beta q_0}} (1+\sqrt{1-e^{-4 \beta q_0}})^2}
 \end{eqnarray}
and where $G(\beta)$ has been defined in (\ref{gbeta}).

In particular, near $T_c$, we have for $T \geq T_c$ the scaling form
\begin{eqnarray}
 Z_{2N}  \opsimeq_{N \hbox{large}, T \to T_c^+} 
\frac{1}{  K_{deloc}^{N} \sqrt{  N}  }
\ \ q \left( \frac{(T-T_c)}{T_c} \sqrt{N C_{\infty}(T_c^-)}  \right)
\label{zfs}
 \end{eqnarray}
leading to the finite size free-energy
\begin{eqnarray}
f_{2N}= \frac{-T \ln Z_{2N} }{2N} \opsimeq_{N \hbox{large}, T \to T_c^+} 
f_{deloc}(T_c)+\frac{T_c}{4N} \ln N 
-\frac{T_c}{2N} \ln q \left( \frac{(T-T_c)}{T_c} \sqrt{N C_{\infty}^-(T_c)}  \right)
\label{freefs}
 \end{eqnarray}

\subsubsection{Specific heat}

In the thermodynamic limit, the specific heat
$C_{\infty}(T)=-T \frac{d^2 f(T)}{d T^2} $ presents a jump at $T_c$
given in equation (\ref{specifinf}).
To compute the finite-size specific heat $C_{2N}(T_c)$ at $T = T_c$,
we use the scaling form (\ref{freefs})

\begin{eqnarray}
C_{2N}(T_c)= && -T \frac{d^2 f_{2N} }{d T^2} 
 \vert_{T=T_c}
= \frac{1}{2} 
\left(\frac{q''(0)}{q(0)}- \left(\frac{q'(0)}{q(0)} \right)^2 \right) C_{\infty}(T_c^-) 
+ \frac{q'(0)}{q(0)} \ \frac{\sqrt{C_{\infty}(T_c^-)}}{\sqrt{N}} +..  \\
= && \frac{4-\pi}{2}  C_{\infty}(T_c^-) -\sqrt{\pi} \frac{\sqrt{C_{\infty}(T_c^-)}}{\sqrt{N}}+..
\label{chaleurfs}
 \end{eqnarray}

Similar relations between $C_{2N}(T_c)$ and  $C_{\infty}(T_c)$ are
found in phase transitions displaying a jump in the specific heat
\cite{Der_Hak_Van,Kit_Sho,Rud_Guo_Jas}.

\subsubsection{ Order parameter and uniform susceptibility}

A possible order parameter at the transition has been mentionned in
equation (\ref{ordparam}), and linked to the probability to be in
fluid $\beta$. We consider here
\begin{eqnarray}
M_{2N} =  \sum_{i=1}^{2N}  \hbox{ sgn} \left[ z(i)-\frac{1}{2} \right] 
 \end{eqnarray}
and the order parameter $m_{2N}=\frac{ < M_{2N} > }{2N}$. Note that
$m_{2N}$ is related to the quantity ${\cal M}_{\beta}$ of
(\ref{ordparam}) by ($m_{2N}=1-2{\cal M}_{\beta}$). Given
the expression of the Hamiltonian (\ref{hamil}), we get
\begin{eqnarray}
m_{\infty} = < \hbox{ sgn} \left[ z(i)-\frac{1}{2}
\right] > =   &&  
 - \partial_h f(T,q_A \to q_A+h,q_B \to q_B-h) \vert_{h=0} \nonumber\\
= && \frac{ \sinh \beta (q_A-q_B) }
 { 2 \sinh \beta q_A \sinh \beta q_B } 
 \opsimeq_{ T \to T_c^-} 1- Cte (T_c-T) +..
 \end{eqnarray}
 In the thermodynamic limit, the corresponding susceptibility reads 
\begin{eqnarray}
\chi_{\infty}= \frac{1}{2TN} \left( < M_{2N}^2 >  - (< M_{2N} > )^2
\right)
= &&  - \partial_h^2 f(T,q_A \to q_A+h,q_B \to q_B-h) \vert_{h=0} \cr
= &&\frac{1}{2 T } \left( \frac{1}{\sinh^2 \beta q_A } +  
\frac{1}{\sinh^2 \beta q_B } \right) 
 \end{eqnarray}
leading, at criticality, to
\begin{eqnarray}
 \chi_{\infty}(T_c^-) = \frac{ 4 \beta_c} {(1- e^{-2 \beta_c q_A})^2}
\label{chiinf}
 \end{eqnarray}

To compute the finite-size behavior of the order parameter $m_{2N}$
and of the susceptibility $\chi_{2N}$, 
we use again the scaling form (\ref{freefs}), but where now $T_c$ is a
function of $h$ defined by the equation for $T_c$ where $q_A \to
q_A+h,q_B \to q_B-h$. Since the calculations exactly parallel the ones 
described above, we only quote the results
\begin{eqnarray}
\label{resultm}
m_{2N}(T_c)=  1 - \frac{\sqrt{2\pi}}{(1- e^{-2 \beta_c q_A})  } \
\frac{1}{\sqrt{2N}} ...
\end{eqnarray}
and
\begin{eqnarray}
\label{resultchi}
\chi_{2N}(T_c)= \frac{4-\pi}{2}  \chi_{\infty}(T_c^-)
\end{eqnarray}

\subsubsection{Numerical results }

We have done numerical calculations using transfer matrix methods with 
both chain ends fixed at the interface. We
chose $q_A=1$ and $q_B=0.5$, yielding a critical temperature
$T_c=1/(\ln (\frac{1+\sqrt{5}}{2}))=2.07809...$ (see equation (\ref{eqtc})).
The corresponding values of the critical specific heat and
susceptibility are $C_{\infty}(T_c^-)=0.11056...$ and
$\chi_{\infty}(T_c^-)=5.03932...$ (see equations (\ref{specifinf}),
(\ref{chiinf})).

Results for $C_{2N}(T)$ and $\chi_{2N}(T)$ in the critical region are
shown in Figures 1 and 2, for chain length up to $2N=18000$ and are in 
quantitative agreement with the above values. We get for instance a
value $C_{2N}(T_c) \simeq 0.04739..$ as compared to the value
$0.047454...$ coming from equation (\ref{chaleurfs}), and
$\chi_{2N}(T_c) \simeq 2.13376...$ as compared to the value $2.16289...$ 
coming from equation (\ref{resultchi}). We have also tested equation
(\ref{resultm}): Figure 3 shows the behaviour of $\sqrt{2N} \
{(1-m_{2N}(T)) \over 2}$ in the critical region. At $T_c$, the
theoretical value coming from equation (\ref{resultm}) is 2.0279... as
compared to the ``experimental'' value 2.0130.. of Figure 3. 

\newpage
\section{The wetting transition }
\label{sec:wetting}
\subsection{Calculation of the grand canonical partition function}
We consider the SOS version of the 2D wetting problem near a
attracting wall at $h=0$. Our presentation will be rather brief, and
we follow the notations of \cite{For_Luc_Nie_Orl} and 
\cite{Der_Hak_Van}. Let us denote by $\{h(i),i=1,2,..,N\}$ the $N$
consecutive heigths 
with the properties $h(i)=0,1,2...$
and $h(i+1)-h(i)=+1,0,-1$. This height model can also be considered as
describing the adsorption of a polymer chain onto the wall. We
further assume that the first height is fixed ($h(i=1)=1$). The
partition function of the model reads
\begin{eqnarray}
\label{partitwett}
Z_{N}=\sum_{\{h(i)\}} 
\exp \left(- \beta J \sum_{i=1}^{N-1} \vert h(i+1)-h(i) \vert
+ \beta u_0 \sum_{i=1}^{N} \delta_{h_i,0}  \right)
\end{eqnarray}

Following closely the steps of the previous section, we first express
the partition function $Z_{N}$ as a function of its end-point ($h(N)=h$), and write
\begin{eqnarray}
Z_{N}=\sum_{h=0}^{+\infty} Z_{N} (h)
\end{eqnarray}
 At this
stage, it is convenient to introduce some notations and relations.

Let ${\cal B}(l)$ represent the partition function for a chain
starting from $h=1$ and arriving at $h=0$ for the first time after $l$
steps. We then denote by ${\cal M}(l-1,h)$ the partition function 
for a chain of $(l-1)$ steps going from $h=1$ to $h \geq 1$ 
in the presence of an absorbing boundary at $h=0$. Setting
$t=e^{-\beta J}$, the method of 
images \cite{Feller} yields
\begin{eqnarray}
  {\cal B}(l) = t \left( R(l-1,0)-R(l-1,2) \right) 
\end{eqnarray}
and
\begin{eqnarray}
  {\cal M}(l-1,h) = R(l-1,h-1)-R(l-1,h+1)
\end{eqnarray}
where $R(l,h)$ denotes the partition function of the chain in the
absence of the wall. We explicitly have
\begin{eqnarray}
 R(l,h)= \sum_{p=\vert h \vert }^l C_l^p C_p^{\frac{p+\vert h \vert }{2}} t^p
\end{eqnarray}
normalized to $\sum_{h=-l}^{+l} R(l,h)= (1+2t)^l$.

We now decompose $Z_{N}(h)$ into free and adsorbed segments as

\begin{eqnarray}
Z_{N}(h)
= && \sum_{p=0}^{+\infty} \sum_{l_1=1}^{\infty} \sum_{l_2=1}^{\infty}
... \sum_{l_{2p}=1}^{\infty} \sum_{l=1}^{\infty}
 \delta(N-( \sum_{i=1}^{2p} l_i+ l) )\\
&& {\cal B}(l_1) ( e^{\beta u_0 l_2} t )
   {\cal B}(l_3)  ( e^{\beta u_0 l_4} t )
 ...
( e^{\beta u_0 l_{2p}}  t ) {\cal M}(l-1,h) \nonumber \ \ \  \ \hbox{for} \ h \geq 1 \nonumber
\\
Z_{N}(0)
 =&&  \sum_{p=0}^{+\infty} \sum_{l_1=1}^{\infty} \sum_{l_2=1}^{\infty}
... \sum_{l_{2p}=1}^{\infty} 
 \delta(N-( \sum_{i=1}^{2p} l_i+ l) )
 {\cal B}(l_1) ( e^{\beta u_0 l_2} t )
 ...
{\cal B}(l_{2p-1}) e^{\beta u_0 l_{2p}}  
\end{eqnarray}

It is again convenient to consider the grand canonical partition function
\begin{eqnarray}
Z(K)=\sum_{N=1}^{+\infty} K^{ N} Z_{N}
\end{eqnarray}
Using
\begin{eqnarray}
{\hat  {\cal B}}(K)=\sum_{l=1}^{+\infty} K^l {\cal B}(l) 
= \frac{1-\sqrt{1-4 \left( \frac{tK}{1-K} \right)^2} }{ 2 \frac{tK}{1-K}}
\end{eqnarray}
we finally have
\begin{eqnarray}
\label{simple1}
Z(K,2h) && = 
\frac{1}{1- {\hat  {\cal B}}(K) \frac{t K e^{\beta u_0} }{1-K e^{\beta u_0}} }
\ \frac{1}{t} \left( \frac{1-K}{tK} {\hat  {\cal B}}(K) -1\right)^h  \\
Z(K,2h+1)&& =
 \frac{1}{1- {\hat  {\cal B}}(K) \frac{t K e^{\beta u_0} }{1-K e^{\beta u_0}} }
\ \frac{1}{t} {\hat  {\cal B}}(K) 
\left( \frac{1-K}{tK} {\hat  {\cal B}}(K)  -1\right)^h
\label{simple2}
\end{eqnarray}

\subsection{Thermodynamics}
Equations (\ref{simple1}) and (\ref{simple2}) give
\begin{eqnarray}
Z(K)=\sum_{h=0}^{+\infty} Z(K,h) =
\frac{1}{1- {\hat  {\cal B}}(K) \frac{t K e^{\beta u_0} }{1-K e^{\beta u_0}} }
\ \frac{1+{\hat  {\cal B}}(K)}{t}
\frac{1}{2-  \frac{1-K}{tK} {\hat  {\cal B}}(K)}
\label{reszz}
\end{eqnarray}
The radius of convergence $K_*(\beta)$ of $Z(K)$ is either given by
the solution of the equation $\left(2-  \frac{1-K}{tK} {\hat  {\cal
B}}(K)=0 \right)$, or by the solution of the equation  $\left(1- {\hat  {\cal B}}(K)
\frac{t K e^{\beta u_0} }{1-K e^{\beta u_0}}=0\right)$. The former
corresponds to a value
\begin{eqnarray}
K_{deloc}= \frac{1}{1+2t} 
\end{eqnarray}
and physically describes the delocalized phase. The latter is given by
\begin{eqnarray}
\label{kstarwett}
K_*(\beta)= \frac{2 e^{-\beta u_0} }
{1+\sqrt{\frac{e^{\beta u_0}-1+4 t^2}{e^{\beta u_0}-1}}}
\end{eqnarray}
The  chain is localized close to the wall as long as $K_*(\beta) < K_{deloc}$,
i.e. $T < T_c=1/\beta_c$ with
\begin{eqnarray}
e^{\beta_c u_0} = \frac{1+2t}{1+t}
\end{eqnarray}
in agreement with references \cite{For_Luc_Nie_Orl,Der_Hak_Van}.

\subsection{Density profile}
We follow equation (\ref{rho}) and define a normalized density profile 
by
\begin{eqnarray}
\rho(h)=\oplimit_{N \to \infty} \frac{Z_{N}(h)}{ Z_{N} }
\end{eqnarray}
Inverting equation (\ref{simple1}, we get
\begin{eqnarray}
\rho(h)=\rho(0) \left[ \frac{1}{2t} \left(
\sqrt{\frac{e^{\beta u_0}-1+4 t^2}{e^{\beta u_0}-1}}-1 \right) \right]^h
\end{eqnarray}
in the localized phase. The characteristic length reads
\begin{eqnarray}
\label{correl}
\xi(T)=-\frac{1}{\ln \left[ \frac{1}{2t} \left(
\sqrt{\frac{e^{\beta u_0}-1+4 t^2}{e^{\beta u_0}-1}}-1 \right) \right] }
\end{eqnarray}
and diverges at the transition as
\begin{eqnarray}
\xi(T) \oppropto_{T \to T_c} \frac{1}{T_c-T}
\end{eqnarray}
in agreement with equation (7.8) of reference \cite{Fis}.
The fraction of adsorbed monomers vanishes as
\begin{eqnarray}
\rho(h=0) \oppropto_{T \to T_c} {T_c-T}
\end{eqnarray}

\subsection{Probability distributions of adsorbed and desorbed segments}

Since the detailed procedure has been given in section
\ref{subsec:probadis}, our presentation will be rather sketchy, and we
only give the main results. Let us denote by A (resp. D) the adsorbed
(resp. desorbed) monomers of the chain. The probability distribution
of desorbed loop lengths will be denoted by $P_D(l)$, and its Laplace
transform by $\hat{P_D}(s)$.
We get
\begin{eqnarray}
\hat{P_D}(s)=\sum_{l=1}^{\infty} e^{-sl} P_D(l) = 
\frac{{\hat  {\cal B}}(K_* e^{-s})}{{\hat  {\cal B}}(K_*)}
\end{eqnarray}
yielding, at criticality, to
\bea
\hat{P_D}(s)\opeq_{T=T_c} {\hat  {\cal B}}(\frac{1}{1+2t}
e^{-s})=1-\sqrt {\frac{1+2t}{t}} \ \sqrt{s} +O(s)
\eea
This corresponds to an algebraic decay of $P_D(l)$ for large $(l)$, as 
in equation (\ref{free}). Similarly, in the region $T \to T_c$
the moments of $P_D(l)$ diverge as
\begin{eqnarray}
\sum_{l=1}^{\infty} l^n P_D(l)  \oppropto_{T \to T_c} (T_c-T) \left( \frac{1}{ (T_c-T)^2} \right)^n 
 \end{eqnarray}
As in the copolymer case, the correlation length along the chain
 scales as $ \left( \frac{1}{ (T_c-T)^2} \right)$, in agreement with
 equation (6.22) of reference \cite{Fis}.
As previously found, the small loops have a finite
weight at $T_c$. One has for instance
\bea
P_D(l=1) \opeq_{T=T_c} \frac{t}{1+2t}
\eea
 As for the probability distribution of the adsorbed segment lengths,
we find an exponential form, even at $T_c$.
\begin{eqnarray}
P_A(l)= (1-K_* e^{\beta u_0} ) (K_* e^{\beta u_0})^{l-1}
\end{eqnarray}
where $K_*$ is given in equation (\ref{kstarwett}).

We conclude this section on wetting by pointing out that a finite size 
relation for the specific heat, quite similar to the one derived in
equation (\ref{chaleurfs}), has been found in reference \cite{Der_Hak_Van}.

\section{Conclusion}
We have presented a grand canonical approach to the localization of a
single polymer chain either at a fluid-fluid interface, or at an attracting
hard wall. This method, which rests on the absence of
self avoidance in the models, provides detailed informations on the 
loop length distribution function. The extension to various disordered situations
\cite{For_Luc_Nie_Orl,Der_Hak_Van,Gar_Hus_Lei_Orl,Ste_Som_Eru,Che,Gan_Bre,Mar_Riv_Tro,Mon,Cha_Joa,Car}}would 
be of great interest.   

\newpage
\global\firstfigfalse
\begin{figure}[htbp]
\inseps{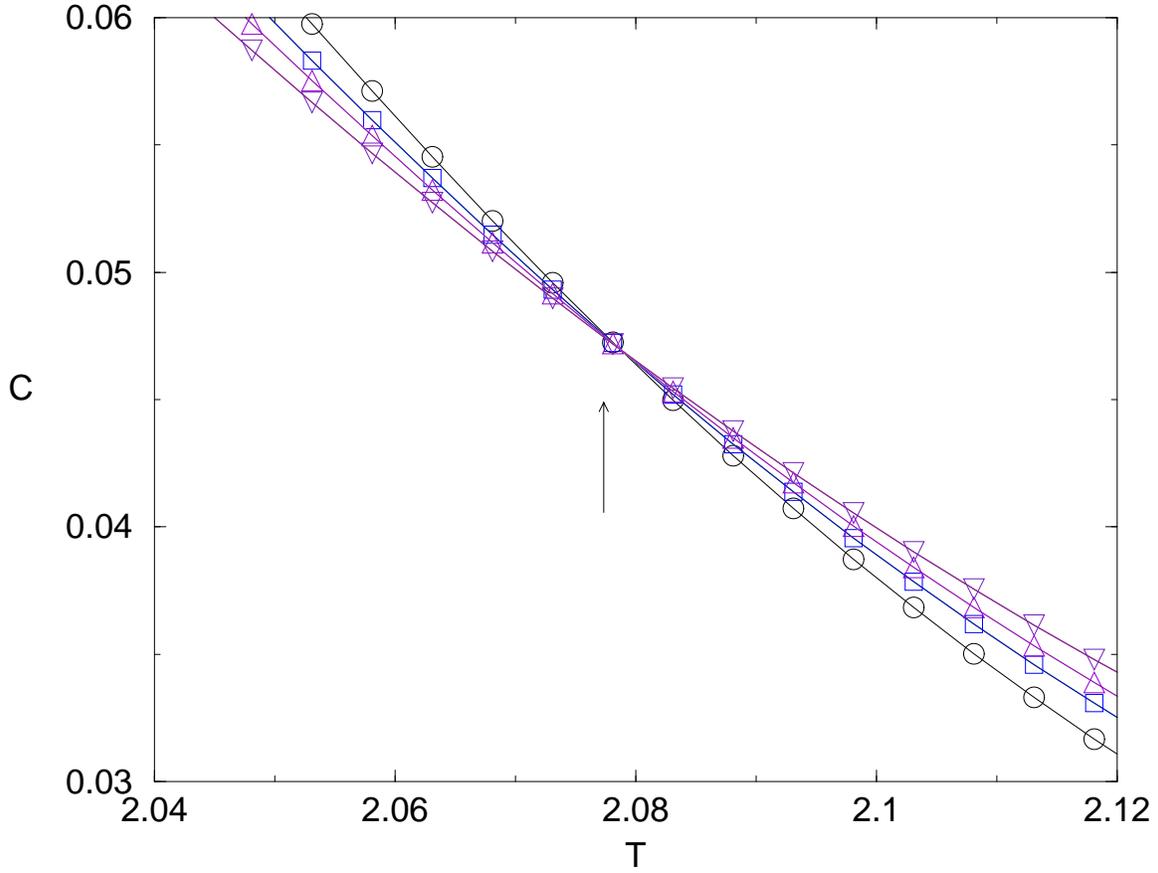}{0.9}
\vskip 8mm
\caption{Specific heat per monomer vs temperature for $q_A=1$ and
$q_B=0.5$ in the critical region for $2N= 18000 \ (\circ)$, $\ 14000 \
(\square)$, $\ 12000 \ (\vartriangle)$, $\ 10000 \
(\triangledown)$. The arrow denotes the critical temperature
$T_c=2.078...$, where equation (\ref{chaleurfs}) predicts $
C(T_c)=0.047454...$.}
\label{figure1}
\end{figure}

\newpage

\begin{figure}
\inseps{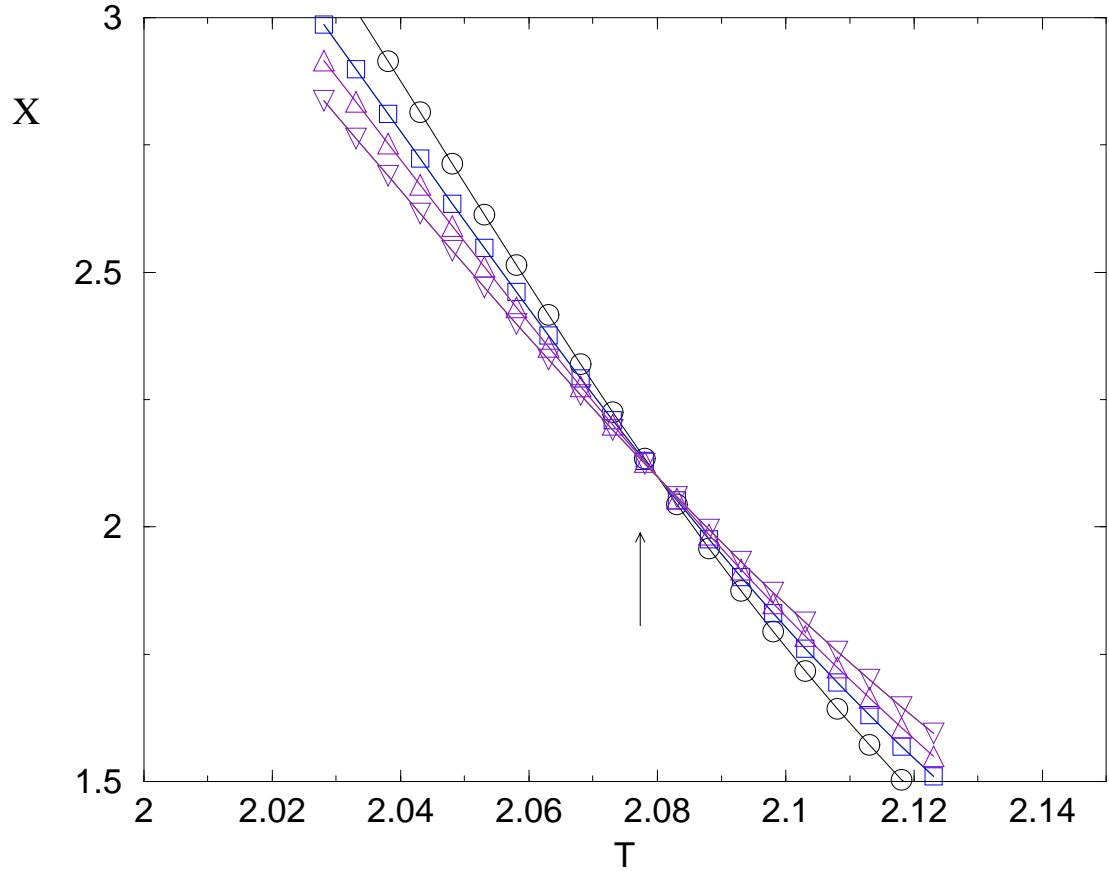}{0.9}
\vskip 8mm
\caption{Susceptibility per monomer vs temperature for $q_A=1$ and
$q_B=0.5$ in the critical region for $2N= 18000 \ (\circ)$, $\ 14000 \
(\square)$, $\ 12000 \ (\vartriangle)$, $\ 10000 \
(\triangledown)$. The arrow denotes the critical temperature
$T_c=2.078...$, where equation (\ref{resultchi}) predicts
$\chi(T_c)=2.16289..$.}
\label{figure2}
\end{figure}

\newpage

\begin{figure}
\inseps{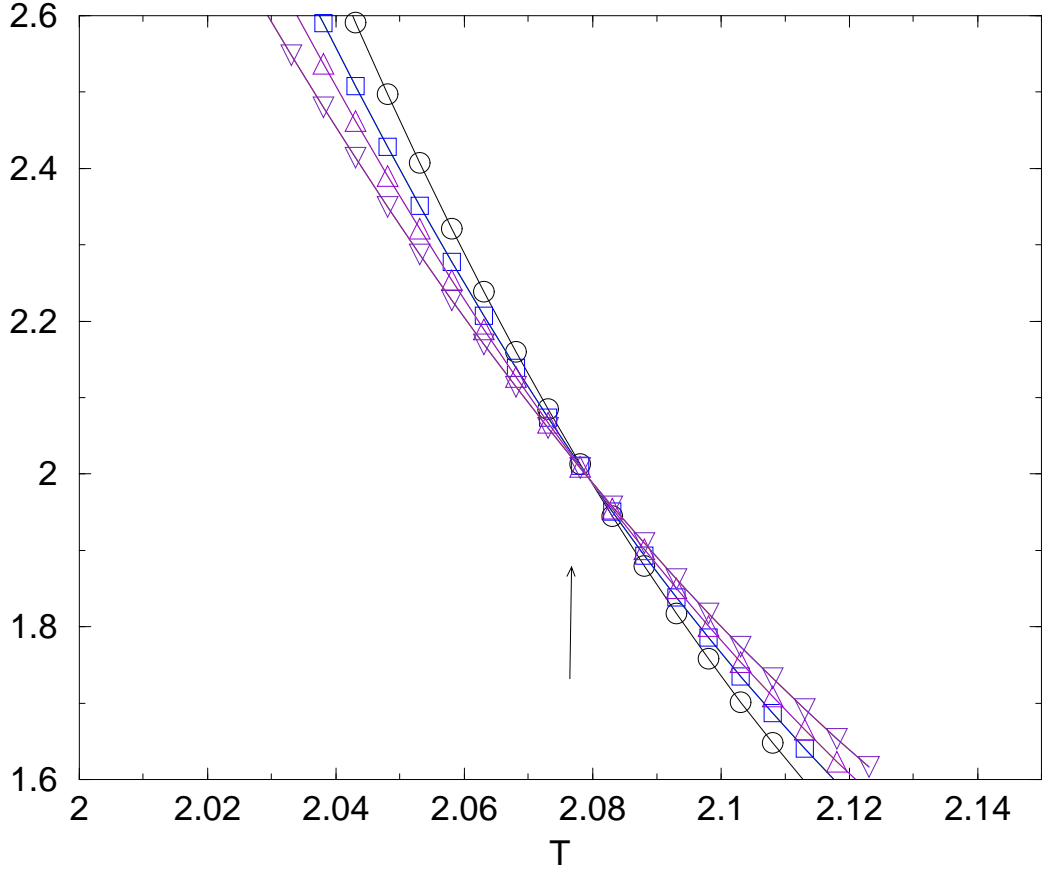}{0.9}
\vskip 8mm
\caption{Test of equation (\ref{resultm}): Plot of $\sqrt{2N} \
{(1-m_{2N}(T)) \over 2}$ vs temperature for $q_A=1$ and
$q_B=0.5$ for $2N= 18000 \ (\circ)$, $\ 14000 \
(\square)$, $\ 12000 \ (\vartriangle)$, $\ 10000 \
(\triangledown)$. The arrow denotes the critical temperature
$T_c=2.078...$, where equation (\ref{resultm}) predicts a value 2.0279..}
\label{figure3}
\end{figure}
\newpage
\begin{appendix}

\section{Transfer matrix approach}

In this appendix, we will briefly present the transfer matrix
calculations for the periodic $AB$ copolymer chain. 
We decompose $Z_N$ according to the position $n$ of the end point of
the chain, ($Z_N= \sum_{n=-\infty}^{+\infty} Z_N(n)$), and using the Hamiltonian 
of equation (\ref{hamil}), we obtain the recursion relations
\begin{eqnarray}
Z_{N+1}(n)= e^{\beta q_{N+1} {\rm sgn} [n-\frac{1}{2}] } 
\left( Z_{N}(n+1) + Z_{N}(n-1) \right)
\end{eqnarray}

It is convenient to introduce the Fourier transforms for the two half-spaces
\begin{eqnarray}
&& { \hat Z}_{N}^+(k)= \sum_{n=1}^{+\infty} e^{ik (n-1) }  Z_N(n) \\
&& { \hat Z}_{N}^-(k)= \sum_{n=-\infty}^{0} e^{ik n }  Z_N(n)
\end{eqnarray}
as well as their inverse
\begin{eqnarray}
&&  Z_N(n)= \int_{0}^{2\pi} \frac{dk} {2\pi} 
e^{-ik (n-1)} { \hat Z}_{N}^+(k) \ \ \ \ \hbox{for} \ \ n \geq 1  \\
&&  Z_N(n)= \int_{0}^{2\pi} \frac{dk} {2\pi} 
e^{-ik n} { \hat Z}_{N}^-(k) \ \ \ \ \hbox{for} \ \ n \leq 0
\label{inversion}
\end{eqnarray}

The recursion relations now read
\begin{eqnarray}
\label{recurr}
&& {\hat Z}_{N+1}^{+}(k)= e^{\beta q_{N+1}  } 
 \left( 2 \cos k {\hat Z}_{N}^{+}(k) 
-  e^{- ik} Z_N(1)+ Z_N(0) \right)  \\
&& {\hat Z}_{N+1}^{-}(k)= e^{-\beta q_{N+1}  } 
 \left( 2 \cos k {\hat Z}_{N}^{-}(k) 
-  e^{ ik} Z_N(0)- Z_N(1) \right)
\end{eqnarray}

To study the thermodynamic limit for our problem, ($q_{2p+1}=q_A \ >0 $
and $q_{2p}=-q_B \ <0$), we look for a stationnary
solution ${\hat Z}_{2N}^{\pm}(k)$ with eigenvalue $\lambda$. We find

\begin{eqnarray}
\label{eigen}
&& {\hat Z}^{+}(k)= 
\frac{1}{ \lambda e^{ \beta (q_B-q_A) } -  4 \cos^2 k   } 
 [ -e^{-ik} Z(2)  
+ e^{ik} Z(0) ]  \\
&& {\hat Z}^{-}(k)= 
\frac{1}{ \lambda e^{- \beta (q_B-q_A) } -  4 \cos^2 k   } 
 [  (e^{2 \beta q_A} -1 -e^{ i 2k}) Z(0) 
+e^{2 \beta q_A} Z(2) ]
\end{eqnarray}

Using the inversion relations (\ref{inversion}), we obtain the
following consistency equations for $Z(2)$ and $Z(0)$
\begin{eqnarray}
&& Z(2)=-I_2^+(\lambda) Z(2) + I_0^+(\lambda) Z(0) \nonumber  \\
&& Z(0)= e^{2 \beta q_A} I_0^-(\lambda) Z(2) +
\left( (e^{2 \beta q_A} -1)  I_0^-(\lambda)-  I_{2}^-(\lambda) \right) Z(0)
\label{consistency}
\end{eqnarray}
whith the notations
\begin{eqnarray}
&& I_p^{\pm} (\lambda)
= \frac{1}{4} J_p \left( y=\frac{\lambda e^{ \pm \beta (q_B-q_A) }}{4} \right) \\
&& J_0(y) = \int_{-\pi}^{+\pi} \frac{dk} {2\pi}
\frac{1}{ y - \cos^2 k   }= \frac{1}{\sqrt{y (y-1)}}  \\
&& J_2(y) = \int_{-\pi}^{+\pi} \frac{dk} {2\pi}
\frac{\cos 2k }{ y - \cos^2 k   }= 
2 \sqrt{ \frac{y}{y-1} }-2-\frac{1}{\sqrt{y (y-1)}} 
\end{eqnarray} 
For a localized eigenstate, both $Z(2)$ and $Z(0)$ are non zero. As a
consequence, the determinant of the system (\ref{consistency})
has to vanish, and this gives an equation for the eigenvalue $\lambda$
of the localized state, which reads in our case ($q_A \geq q_B$)
\bea
\lambda={1 \over K_*(\beta)}
\eea
where $K_*(\beta)$ is given in equation (\ref{kstar}). The free
energy is then given by
\begin{eqnarray}
f(T) = \lim_{N \to \infty} \frac{- \ln Z_{2N} }{ \beta 2 N}
= - \frac{ \ln \lambda }{ 2 \beta}
\end{eqnarray}
in agreement with equation (\ref{free ener}).

As a final remark on this approach, we mention how one may find the polymer
density profile  $\rho(2n)$ in the localized phase.
The Fourier transforms of $\rho(2n)$ on the two-half spaces
are indeed directly related to the localized eigenvector $({\hat
Z}^{+}(k))$ given in (\ref{eigen}) by

\begin{eqnarray}
&& \sum_{n=0}^{+\infty} e^{ik (2n-1) } \rho(2n)
= \frac{{\hat Z}^{+}(k)}{{\hat Z}^{+}(0)+{\hat Z}^{-}(0)}
 \\
&& \sum_{n=-\infty}^{0} e^{ik 2n } \rho(2n)
= \frac{{\hat Z}^{-}(k)}{{\hat Z}^{+}(0)+{\hat Z}^{-}(0)}
\end{eqnarray}
in agreement with the result (\ref{pstar}). All the results of section 
\ref{sec:periodic} can be thus recovered, except for the results of
the loop length distribution function of section \ref{subsec:probadis}.
\end{appendix}

\end{document}